\documentclass[showpacs,aps,prd]{revtex4}
\input epsf

\begin{document}

\title{Exploring a $\Sigma_{c}\bar{D}$ state: with focus on $P_{c}(4312)^{+}$}
\author{Jian-Rong Zhang}
\affiliation{Department of Physics, College of Liberal Arts and Sciences, National University of Defense Technology,
Changsha 410073, Hunan, People's Republic of China}

%\date{\today}

%%%%%%%%%%%%%%%%%%%%%%%%%%%%%%%%%%%%%%%%%%%%%%%%%%%%%%%%%%%%%%%%%%%%%
\begin{abstract}
Stimulated by the new discovery of $P_{c}(4312)^{+}$ by LHCb Collaboration, we
endeavor to perform the study of $P_{c}(4312)^{+}$ as a
$\Sigma_{c}\bar{D}$ state
in the framework of QCD sum rules.
Taking into account the results from two sum rules,
a conservative mass range $4.07\sim4.97~\mbox{GeV}$
is presented for the $\Sigma_{c}\bar{D}$ hadronic system,
which agrees with the experimental data of $P_{c}(4312)^{+}$ and could support
its interpretation as a $\Sigma_{c}\bar{D}$ state.
\end{abstract}
\pacs {11.55.Hx, 12.38.Lg, 12.39.Mk}\maketitle

Keywords: QCD sum rules; $P_{c}(4312)^{+}$

\section{Introduction}\label{sec1}
Very recently, LHCb Collaboration reported the discovery of a narrow state
$P_{c}(4312)^{+}$ with a
statistical significance of $7.3\sigma$ in a data sample of $\Lambda_{b}^{0}\rightarrow J/\psi p K^{-}$ decays \cite{LHCb}.
Moreover, $P_{c}(4450)^{+}$ formerly announced by LHCb is confirmed and
observed to consist of two narrow overlapping peaks, $P_{c}(4440)^{+}$ and $P_{c}(4457)^{+}$.
Soon after the LHCb's new observation,
many works \cite{th-1,th-2,th-3,th-4,th-5,th-6,th-7,th-8,th-9,th-10,th-11,th-12,th-13,th-14,th-15,th-16} have been promptly triggered.
Among these new experimental results,
the most exciting point should attribute to the freshly discovered
$P_{c}(4312)^{+}$.
After all, there already have existed plenty of researches  on $P_{c}(4450)^{+}$ \cite{others} (one can
also see a
recent review e.g. \cite{Penta-overview}).
Besides, $P_{c}(4312)^{+}$ is narrow
and below the $\Sigma_{c}^{+}\bar{D}^{0}$ threshold within a plausible hadron-hadron
binding energy, hence it provides the strongest experimental evidence to date for the existence
of a $\Sigma_{c}\bar{D}$ bound state \cite{LHCb}.
Meanwhile, some different opinion has also appeared in Ref. \cite{th-15},
in which the authors could find evidence for the attractive effect
of the $\Sigma_{c}^{+}\bar{D}^{0}$ channel, however not strong enough to form a bound state
and they infer that the $P_{c}(4312)^{+}$ peak is
more likely to be a virtual (unbound) state instead.
Whether or no, to realize the nature of $P_{c}(4312)^{+}$, it certainly requires more theoretical scrutiny.

In this work, we focus all our attention on the
newly discovered $P_{c}(4312)^{+}$ and would
investigate the possibility of $P_{c}(4312)^{+}$ being a $\Sigma_{c}\bar{D}$ state,
if the $\Sigma_{c}\bar{D}$ state does exist.
While studying a baryon-meson state, one inevitably
has to confront and treat nonperturbative QCD problem.
As one reliable method for evaluating
nonperturbative effects, the QCD sum rule \cite{svzsum} is an
analytic formalism firmly established on QCD theory and
has been successfully applied
to different hadronic systems \cite{overview1,overview2,overview3,reinders,overview4}.
As a matter of fact, there have appeared some related works on
these $P_{c}$ hadrons basing on baryon-meson configuration QCD sum rules \cite{th-1,Peta-SR1,Peta-SR2,Peta-SR3,Peta-SR4}.
In QCD sum rule analysis, it is of great importance to carefully inspect
both the operator product expansion (OPE) convergence and
the pole dominance in order to ensure the extracted result authentic.
In practice, one could note that some condensate may
play an important role in some multiquark cases \cite{Zs0,Zs1,Zs2,Zs},
which causes that it is of difficulty to find
conventional work windows.
Specially for the four-quark condensate, a general factorization $\langle\bar{q}q\bar{q}q\rangle=\varrho\langle\bar{q}q\rangle^{2}$ has been hotly discussed \cite{A,B}, where $\varrho$
is a constant, which may be equal to $1$, to $2$, or be smaller than $1$.
Moreover, the factorization parameter $\varrho$ could be about $3\sim4$ \cite{C}. Compromisingly,
the parameter $\varrho$ is taken as $2$ in this work.

The rest paper is organized as follows. In Sec. \ref{sec2}, $P_{c}(4312)^{+}$ is
studied as a $\Sigma_{c}\bar{D}$ state through the QCD sum
rule approach. Numerical analysis and discussions
are given in Sec. \ref{sec3}.
The last part is a brief
summary.
%%%%%%%%%%%%%%%%%%%%%%%%%%%%%%%%%%%%%%%%%%%%%%%%%%%%%%%%%%%%%%%%%%%
\section{QCD sum rule study of $P_{c}(4312)^{+}$ as a $\Sigma_{c}\bar{D}$ state}\label{sec2}
Mass sum rules for a $\Sigma_{c}\bar{D}$ state can be derived from the two-point correlator
\begin{eqnarray}\label{correlator}
\Pi(q^{2})=i\int
d^{4}x\mbox{e}^{iq.x}\langle0|T[j(x)\overline{j}(0)]|0\rangle.
\end{eqnarray}
To represent the $\Sigma_{c}\bar{D}$ state,
one can construct its interpolating current $j$ from
baryon-meson type of fields
adopting currents for the heavy baryon \cite{baryon-current} and for the heavy
meson \cite{reinders}.
Concretely, the current can be written as
\begin{eqnarray}
j=\epsilon_{abe}(q_{a}^{T}C\gamma_{\mu}q_{b})\gamma^{\mu}\gamma_{5}c_{e}\bar{c}_{f}i\gamma_{5}q_{f}.\nonumber
\end{eqnarray}
Here $q$ could be the light $u$ or $d$ quark, $c$ denotes the heavy charm quark,
$T$ means matrix transposition, $C$ is the charge conjugation matrix,
and the subscript $a$, $b$, $e$, and $f$ are color indices.

Lorentz covariance implies that the
two-point correlator (\ref{correlator}) has the
general form
\begin{eqnarray}
\Pi(q^{2})=\Pi_{1}(q^{2})+\rlap/q\Pi_{2}(q^{2}).
\end{eqnarray}
In phenomenology, it can be
expressed as
\begin{eqnarray}\label{pole-model}
\Pi(q^{2})=\lambda^{2}_H\frac{M_{H}+\rlap/q}{M_{H}^{2}-q^{2}}+\frac{1}{\pi}\int_{s_{0}}
^{\infty}ds\frac{\mbox{Im}\Pi_{1}^{\mbox{phen}}+\rlap/q\mbox{Im}\Pi_{2}^{\mbox{phen}}}{s-q^{2}}+...
\end{eqnarray}
where $M_{H}$ is the hadron's mass, and
$\lambda_{H}$ denotes the coupling of the current to the hadron
$\langle0|j|H\rangle=\lambda_{H}u(p,s)$. In the OPE side,
one can write the correlator as
\begin{eqnarray}
\Pi(q^{2})=\int_{4m_{c}^{2}}^{\infty}ds\frac{\rho_{1}}{s-q^{2}}+\rlap/q\int_{4m_{c}^{2}}^{\infty}ds\frac{\rho_{2}}{s-q^{2}}+...
\end{eqnarray}
where spectral densities are $\rho_{i}=\frac{1}{\pi}\mbox{Im}\Pi_{i}^{\mbox{OPE}}$, with $i=1,2$.
After equating the two expressions, applying quark-hadron duality, and making a Borel transform, the sum
rules are
\begin{eqnarray}\label{sumrule1}
\lambda^{2}_HM_{H}e^{-M_{H}^{2}/M^{2}}&=&\int_{4m_{c}^{2}}^{s_{0}}ds\rho_{1}e^{-s/M^{2}},
\end{eqnarray}
and
\begin{eqnarray}\label{sumrule2}
\lambda^{2}_He^{-M_{H}^{2}/M^{2}}&=&\int_{4m_{c}^{2}}^{s_{0}}ds\rho_{2}e^{-s/M^{2}},
\end{eqnarray}
where $M^2$ indicates the Borel parameter.
Taking the derivative of Eq. (\ref{sumrule1}) or (\ref{sumrule2}) with respect to $1/M^{2}$ and dividing the equation itself,
one can obtain mass sum rules
\begin{eqnarray}\label{sum rule m}
M_{H}^{2}&=&\int_{4m_{c}^{2}}^{s_{0}}ds\rho_{1}s
e^{-s/M^{2}}/
\int_{4m_{c}^{2}}^{s_{0}}ds\rho_{1}e^{-s/M^{2}},
\end{eqnarray}
and
\begin{eqnarray}\label{sum rule q}
M_{H}^{2}&=&\int_{4m_{c}^{2}}^{s_{0}}ds\rho_{2}s
e^{-s/M^{2}}/
\int_{4m_{c}^{2}}^{s_{0}}ds\rho_{2}e^{-s/M^{2}}.
\end{eqnarray}

In the deriving of spectral densities, one can
utilize the similar techniques as Refs. e.g. \cite{overview4,Zhang}.
The heavy-quark propagator in momentum-space \cite{reinders}
can be used to keep the heavy-quark mass finite, and
the correlator's light-quark part can be obtained in the coordinate space,
which is then
Fourier-transformed to the $D$ dimension momentum space.
The resulting light-quark part is combined with the heavy-quark part
before it is dimensionally regularized at $D=4$.
As follows, we concretely present spectral densities $\rho_{i}$
deduced from $\Pi_{i}(q^{2})$ and put them forward to further numerical analysis, with
\begin{eqnarray}
\rho_{i}=\rho_{i}^{\mbox{pert}}+\rho_{i}^{\langle\bar{q}q\rangle}+\rho_{i}^{\langle
g^{2}G^{2}\rangle}+\rho_{i}^{\langle
g\bar{q}\sigma\cdot G q\rangle}+\rho_{i}^{\langle\bar{q}q\rangle^{2}}+\rho_{i}^{\langle
g^{3}G^{3}\rangle}+\rho_{i}^{\langle\bar{q}q\rangle\langle
g^{2}G^{2}\rangle}+\rho_{i}^{\langle\bar{q}q\rangle\langle g\bar{q}\sigma\cdot G q\rangle}\nonumber
\end{eqnarray}
up to dimension $8$. In detail,
\begin{eqnarray}
\rho_{1}^{\mbox{pert}}&=&-\frac{1}{5\cdot2^{14}\pi^{8}}m_{c}\int_{\alpha_{min}}^{\alpha_{max}}\frac{d\alpha}{\alpha^{5}}\int_{\beta_{min}}^{1-\alpha}\frac{d\beta}{\beta^{4}}(1-\alpha-\beta)^{3}[(\alpha+\beta)m_{c}^{2}-\alpha\beta s]^{5},\nonumber\\
\rho_{1}^{\langle\bar{q}q\rangle}&=&\frac{\langle\bar{q}q\rangle}{2^{10}\pi^{6}}m_{c}^{2}\int_{\alpha_{min}}^{\alpha_{max}}\frac{d\alpha}{\alpha^{3}}\int_{\beta_{min}}^{1-\alpha}\frac{d\beta}{\beta^{3}}(1-\alpha-\beta)^{2}[(\alpha+\beta)m_{c}^{2}-\alpha\beta s]^{3},\nonumber\\
\rho_{1}^{\langle g^{2}G^{2}\rangle}&=&-\frac{\langle
g^{2}G^{2}\rangle}{3\cdot2^{15}\pi^{8}}m_{c}\int_{\alpha_{min}}^{\alpha_{max}}\frac{d\alpha}{\alpha^{5}}\int_{\beta_{min}}^{1-\alpha}\frac{d\beta}{\beta^{4}}(1-\alpha-\beta)^{3}[(\alpha+\beta)m_{c}^{2}-\alpha\beta s]^{2}[(\alpha+\beta)(\alpha^{2}-\alpha\beta\nonumber\\
&+&2\beta^{2})m_{c}^{2}-\alpha\beta^{3}s],\nonumber\\
\rho_{1}^{\langle g\bar{q}\sigma\cdot G q\rangle}&=&\frac{3\langle
g\bar{q}\sigma\cdot G
q\rangle}{2^{11}\pi^{6}}m_{c}^{2}\int_{\alpha_{min}}^{\alpha_{max}}\frac{d\alpha}{\alpha^{2}}\int_{\beta_{min}}^{1-\alpha}\frac{d\beta}{\beta^{2}}(1-\alpha-\beta)[(\alpha+\beta)m_{c}^{2}-\alpha\beta s]^{2},\nonumber\\
\rho_{1}^{\langle\bar{q}q\rangle^{2}}&=&\frac{\varrho\langle\bar{q}q\rangle^{2}}{2^{6}\pi^{4}}m_{c}\int_{\alpha_{min}}^{\alpha_{max}}\frac{d\alpha}{\alpha^{2}}\int_{\beta_{min}}^{1-\alpha}\frac{d\beta}{\beta}[(\alpha+\beta)m_{c}^{2}-\alpha\beta s]^{2},\nonumber\\
\rho_{1}^{\langle g^{3}G^{3}\rangle}&=&-\frac{\langle
g^{3}G^{3}\rangle}{3\cdot2^{17}\pi^{8}}m_{c}\int_{\alpha_{min}}^{\alpha_{max}}\frac{d\alpha}{\alpha^{5}}\int_{\beta_{min}}^{1-\alpha}\frac{d\beta}{\beta^{4}}(1-\alpha-\beta)^{3}[(\alpha+\beta)m_{c}^{2}-\alpha\beta s]\Big\{(\alpha^{3}+6\beta^{3})[(\alpha+\beta)m_{c}^{2}\nonumber\\&-&\alpha\beta s]+4m_{c}^{2}(\alpha^{4}+\beta^{4})\Big\},\nonumber\\
\rho_{1}^{\langle\bar{q}q\rangle\langle
g^{2}G^{2}\rangle}&=&\frac{\langle\bar{q}q\rangle\langle
g^{2}G^{2}\rangle}{3\cdot2^{12}\pi^{6}}m_{c}^{2}\int_{\alpha_{min}}^{\alpha_{max}}\frac{d\alpha}{\alpha^{3}}\int_{\beta_{min}}^{1-\alpha}\frac{d\beta}{\beta^{3}}\Big\{[\alpha^{2}\beta^{2}+3(\alpha^{2}+\beta^{2})(1-\alpha-\beta)^{2}][(\alpha+\beta)m_{c}^{2}-\alpha\beta s]\nonumber\\
&+&(\alpha^{3}+\beta^{3})(1-\alpha-\beta)^{2}m_{c}^{2}\Big\},\nonumber\\
\rho_{1}^{\langle\bar{q}q\rangle\langle g\bar{q}\sigma\cdot G q\rangle}&=&\frac{\langle\bar{q}q\rangle\langle g\bar{q}\sigma\cdot G q\rangle}{2^{6}\pi^{4}}m_{c}\int_{\alpha_{min}}^{\alpha_{max}}\frac{d\alpha}{\alpha}[m_{c}^{2}-\alpha(1-\alpha)s],\nonumber\\
\rho_{2}^{\mbox{pert}}&=&-\frac{1}{5\cdot2^{13}\pi^{8}}\int_{\alpha_{min}}^{\alpha_{max}}\frac{d\alpha}{\alpha^{4}}\int_{\beta_{min}}^{1-\alpha}\frac{d\beta}{\beta^{4}}(1-\alpha-\beta)^{3}[(\alpha+\beta)m_{c}^{2}-\alpha\beta s]^{5},\nonumber\\
\rho_{2}^{\langle\bar{q}q\rangle}&=&\frac{\langle\bar{q}q\rangle}{2^{9}\pi^{6}}m_{c}\int_{\alpha_{min}}^{\alpha_{max}}\frac{d\alpha}{\alpha^{2}}\int_{\beta_{min}}^{1-\alpha}\frac{d\beta}{\beta^{3}}(1-\alpha-\beta)^{2}[(\alpha+\beta)m_{c}^{2}-\alpha\beta s]^{3},\nonumber\\
\rho_{2}^{\langle g^{2}G^{2}\rangle}&=&-\frac{\langle
g^{2}G^{2}\rangle}{3\cdot2^{14}\pi^{8}}m_{c}^{2}\int_{\alpha_{min}}^{\alpha_{max}}\frac{d\alpha}{\alpha^{4}}\int_{\beta_{min}}^{1-\alpha}\frac{d\beta}{\beta^{4}}(\alpha^{3}+\beta^{3})(1-\alpha-\beta)^{3}[(\alpha+\beta)m_{c}^{2}-\alpha\beta s]^{2},\nonumber\\
\rho_{2}^{\langle g\bar{q}\sigma\cdot G q\rangle}&=&\frac{3\langle
g\bar{q}\sigma\cdot G
q\rangle}{2^{10}\pi^{6}}m_{c}\int_{\alpha_{min}}^{\alpha_{max}}\frac{d\alpha}{\alpha}\int_{\beta_{min}}^{1-\alpha}\frac{d\beta}{\beta^{2}}(1-\alpha-\beta)[(\alpha+\beta)m_{c}^{2}-\alpha\beta s]^{2},\nonumber\\
\rho_{2}^{\langle\bar{q}q\rangle^{2}}&=&\frac{\varrho\langle\bar{q}q\rangle^{2}}{2^{7}\pi^{4}}\int_{\alpha_{min}}^{\alpha_{max}}\frac{d\alpha}{\alpha}\int_{\beta_{min}}^{1-\alpha}\frac{d\beta}{\beta}[(\alpha+\beta)m_{c}^{2}-\alpha\beta s]^{2},\nonumber\\
\rho_{2}^{\langle g^{3}G^{3}\rangle}&=&-\frac{\langle
g^{3}G^{3}\rangle}{3\cdot2^{16}\pi^{8}}\int_{\alpha_{min}}^{\alpha_{max}}\frac{d\alpha}{\alpha^{4}}\int_{\beta_{min}}^{1-\alpha}\frac{d\beta}{\beta^{4}}(1-\alpha-\beta)^{3}[(\alpha+\beta)m_{c}^{2}-\alpha\beta s]\Big\{\beta^{3}[(\alpha+5\beta)m_{c}^{2}-\alpha\beta s]\nonumber\\&+&\alpha^{3}[(5\alpha+\beta)m_{c}^{2}-\alpha\beta s]\Big\},\nonumber\\
\rho_{2}^{\langle\bar{q}q\rangle\langle
g^{2}G^{2}\rangle}&=&\frac{\langle\bar{q}q\rangle\langle
g^{2}G^{2}\rangle}{3\cdot2^{11}\pi^{6}}m_{c}\int_{\alpha_{min}}^{\alpha_{max}}\frac{d\alpha}{\alpha^{2}}\int_{\beta_{min}}^{1-\alpha}\frac{d\beta}{\beta^{3}}\Big\{(\alpha+\beta)(4\alpha^{2}-\alpha\beta+\beta^{2})(1-\alpha-\beta)^{2}m_{c}^{2}\nonumber\\
&-&\alpha^{3}\beta [\beta^{2}+3(1-\alpha-\beta)^{2}]s\Big\},\nonumber
\end{eqnarray}
and
\begin{eqnarray}
\rho_{2}^{\langle\bar{q}q\rangle\langle g\bar{q}\sigma\cdot G q\rangle}&=&\frac{\langle\bar{q}q\rangle\langle g\bar{q}\sigma\cdot G q\rangle}{2^{7}\pi^{4}}\int_{\alpha_{min}}^{\alpha_{max}}d\alpha[m_{c}^{2}-\alpha(1-\alpha)s],\nonumber
\end{eqnarray}
in which the general $\langle\bar{q}q\bar{q}q\rangle=\varrho\langle\bar{q}q\rangle^{2}$ factorization
has been used.
The integration limits are
$\alpha_{min}=\Big(1-\sqrt{1-4m_{c}^{2}/s}\Big)/2$,
$\alpha_{max}=\Big(1+\sqrt{1-4m_{c}^{2}/s}\Big)/2$, and
$\beta_{min}=\alpha m_{c}^{2}/(s\alpha-m_{c}^{2})$.
Those condensates higher than dimension $8$
are not involved here,
as one could expect that kind of high dimension contributions
may not radically influence the OPE's character \cite{Zhang1,Zhang2}.

\section{numerical analysis and discussions}\label{sec3}
In this part, we firstly perform the numerical analysis of sum rule (\ref{sum rule q})
to extract the value of $M_{H}$,
and take $m_{c}$
as the running charm quark mass $1.275_{-0.035}^{+0.025}~\mbox{GeV}$ \cite{PDG}
along with other input parameters as
$\langle\bar{q}q\rangle=-(0.24\pm0.01)^{3}~\mbox{GeV}^{3}$,
$\langle
g\bar{q}\sigma\cdot G q\rangle=m_{0}^{2}~\langle\bar{q}q\rangle$,
$m_{0}^{2}=0.8\pm0.1~\mbox{GeV}^{2}$, $\langle
g^{2}G^{2}\rangle=0.88\pm0.25~\mbox{GeV}^{4}$, and $\langle
g^{3}G^{3}\rangle=0.58\pm0.18~\mbox{GeV}^{6}$ \cite{svzsum,overview2}.
Steering a middle course, the factorization parameter $\varrho$ is set to be $2$.
According to a standard procedure,
both the OPE convergence and the pole dominance should be considered
to find appropriate work windows for the threshold $\sqrt{s_{0}}$ and the Borel
parameter:
the lower bound of $M^{2}$ is gained by analyzing the OPE
convergence, and
the upper one is obtained by
viewing that the pole contribution should be larger
than QCD continuum contribution.
Besides, $\sqrt{s_{0}}$ characterizes the
beginning of continuum states and should not be taken at will.
It is correlated to
the next excited state energy
and empirically
$400\sim600~\mbox{MeV}$ above the eventually achieved
value $M_{H}$.

In FIG. 1, the relative contributions of various OPE in sum rule
(\ref{sumrule2})
are compared as a function of $M^2$
for the $\Sigma_{c}\bar{D}$ state.
Visually, there
four main condensate contributions could play an important role on the OPE side, i.e. the two-quark condensate $\langle\bar{q}q\rangle$,
the mixed condensate $\langle
g\bar{q}\sigma\cdot G q\rangle$, the four-quark condensate $\langle\bar{q}q\rangle^{2}$,
and the $\langle\bar{q}q\rangle\langle
g\bar{q}\sigma\cdot G q\rangle$ condensate.
The direct consequence is that it is not easy to find the standard Borel window,
in which the low dimension condensate contribution should be bigger than the high
dimension one. To say the least,
these four main condensates
could cancel each other out
to some extent.
In this way, the perturbative term still plays an important role on the OPE side
and the OPE's convergence could be under control
at the relatively low value of $M^{2}$.
Thus, the lower bound of $M^{2}$ is taken as $2.0~\mbox{GeV}^{2}$ for
the sum rule
(\ref{sumrule2}).

\begin{figure}[htb!]
\centerline{\epsfysize=7.0truecm\epsfbox{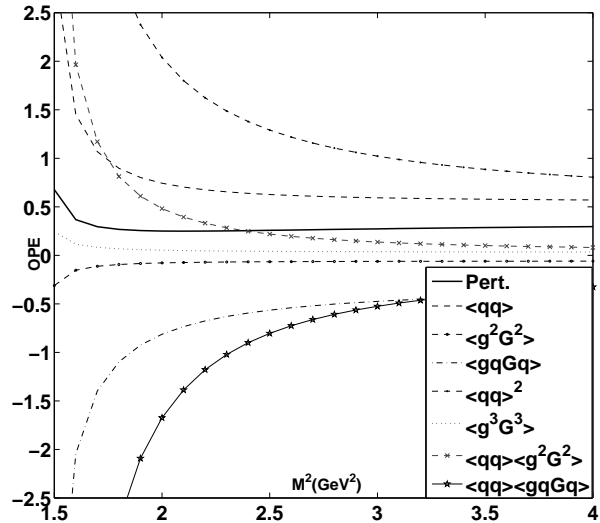}}
\caption{The relative contributions of various OPE as a function of $M^2$ in sum rule
(\ref{sumrule2}) for $\sqrt{s_{0}}=4.8~\mbox{GeV}$ for $\Sigma_{c}\bar{D}$.}
\end{figure}

Phenomenologically,
a comparison between pole contribution and
continuum contribution of sum rule (\ref{sumrule2}) for
$\sqrt{s_{0}}=4.8~\mbox{GeV}$ is shown in FIG. 2, which manifests that the relative pole
contribution is about $50\%$ at $M^{2}=2.7~\mbox{GeV}^{2}$
and decreases with $M^{2}$. In a similar way, the upper bounds of Borel parameters are
$M^{2}=2.6~\mbox{GeV}^{2}$ for
$\sqrt{s_0}=4.7~\mbox{GeV}$ and $M^{2}=2.9~\mbox{GeV}^{2}$ for
$\sqrt{s_0}=4.9~\mbox{GeV}$.
Thus, Borel windows are taken as
$2.0\sim2.6~\mbox{GeV}^{2}$ for $\sqrt{s_0}=4.7~\mbox{GeV}$,
$2.0\sim2.7~\mbox{GeV}^{2}$ for
$\sqrt{s_0}=4.8~\mbox{GeV}$, and
$2.0\sim2.9~\mbox{GeV}^{2}$ for $\sqrt{s_0}=4.9~\mbox{GeV}$.
The mass $M_{H}$ of $\Sigma_{c}\bar{D}$
is shown in FIG. 3 as a function of $M^2$ from sum rule (\ref{sum rule q}).
In the chosen work windows,
$M_{H}$ is calculated to be $4.35\pm0.07~\mbox{GeV}$.
Furthermore, in view of the uncertainty due to the variation of quark masses and
condensates, we have
$4.35\pm0.07^{+0.55}_{-0.21}~\mbox{GeV}$ (the
first error is resulted from the variation of $\sqrt{s_{0}}$
and $M^{2}$, and the second error reflects the uncertainty rooting in the variation of
QCD parameters) or briefly $4.35^{+0.62}_{-0.28}~\mbox{GeV}$
for $\Sigma_{c}\bar{D}$.

\begin{figure}[htb!]
\centerline{\epsfysize=7.0truecm\epsfbox{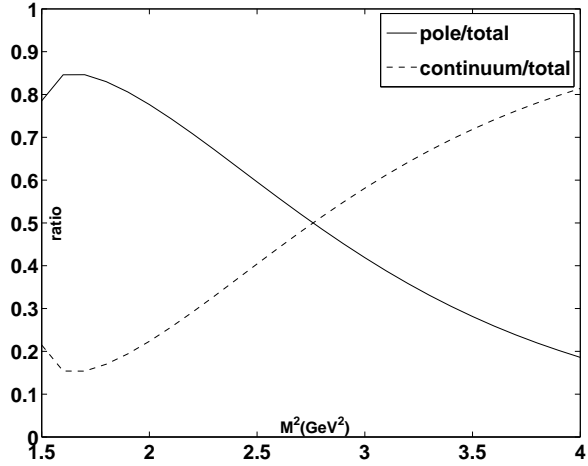}}
\caption{The phenomenological contribution in sum rule
(\ref{sumrule2}) for $\sqrt{s_{0}}=4.8~\mbox{GeV}$ for $\Sigma_{c}\bar{D}$.
The solid line is the relative pole contribution (the pole
contribution divided by the total, pole plus continuum contribution)
as a function of $M^2$ and the dashed line is the relative continuum
contribution.}
\end{figure}

\begin{figure}[htb!]
\centerline{\epsfysize=7.0truecm
\epsfbox{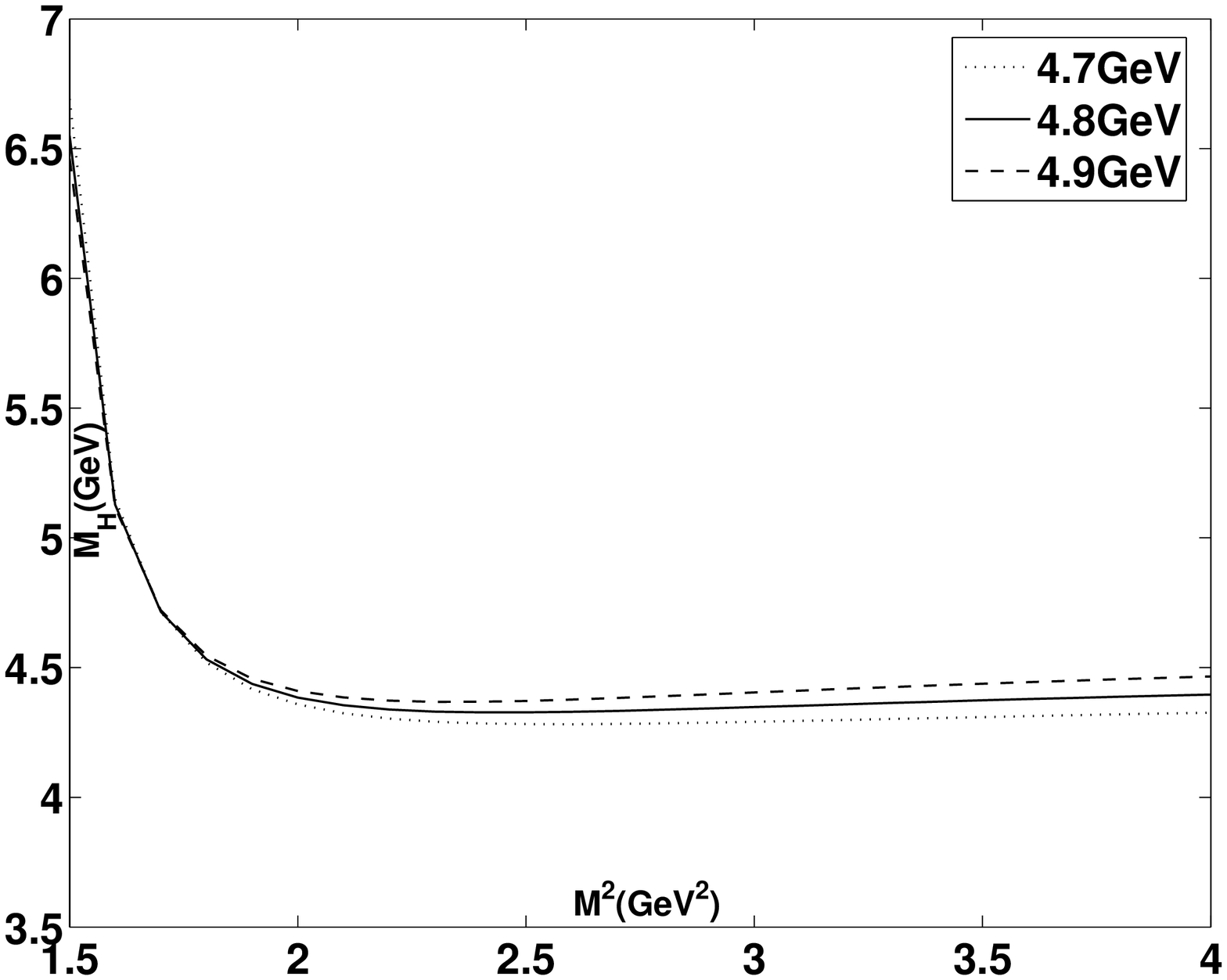}}\caption{
The mass of $\Sigma_{c}\bar{D}$ state as
a function of $M^2$ from sum rule (\ref{sum rule q}). The continuum
thresholds are taken as $\sqrt{s_0}=4.7\sim4.9~\mbox{GeV}$. The
ranges of $M^{2}$ are $2.0\sim2.6~\mbox{GeV}^{2}$ for
$\sqrt{s_0}=4.7~\mbox{GeV}$, $2.0\sim2.7~\mbox{GeV}^{2}$ for
$\sqrt{s_0}=4.8~\mbox{GeV}$, and $2.0\sim2.9~\mbox{GeV}^{2}$ for
$\sqrt{s_0}=4.9~\mbox{GeV}$.}
\end{figure}

Furthermore, one could put forward the numerical analysis of sum rule (\ref{sum rule m}) analogously.
In FIG. 4, the relative contributions of various OPE in sum rule
(\ref{sumrule1}) are shown
 as a function of $M^2$
 for $\sqrt{s_{0}}=4.8~\mbox{GeV}$.
Similarly, four main condensates (i.e. $\langle\bar{q}q\rangle$,
$\langle g\bar{q}\sigma\cdot G q\rangle$, $\langle\bar{q}q\rangle^{2}$,
and $\langle\bar{q}q\rangle\langle g\bar{q}\sigma\cdot G q\rangle$)
could cancel each other out
to some extent. For the sum rule
(\ref{sumrule1}), the lower bound of $M^{2}$ is taken as $2.2~\mbox{GeV}^{2}$
at which the OPE's convergence could still be controllable.
In FIG. 5,
a comparison between pole and
continuum contribution of sum rule (\ref{sumrule1}) is shown  for
$\sqrt{s_{0}}=4.8~\mbox{GeV}$, which indicates that the relative pole
contribution is about $50\%$ at $M^{2}=2.9~\mbox{GeV}^{2}$
and decreases with $M^{2}$.
Thereby, the ranges of $M^{2}$ are fixed as $2.2\sim2.9~\mbox{GeV}^{2}$ for
$\sqrt{s_0}=4.7~\mbox{GeV}$, $2.2\sim3.1~\mbox{GeV}^{2}$ for
$\sqrt{s_0}=4.8~\mbox{GeV}$, and $2.2\sim3.2~\mbox{GeV}^{2}$ for
$\sqrt{s_0}=4.9~\mbox{GeV}$.
The mass of $\Sigma_{c}\bar{D}$ state is shown in FIG. 6 as
a function of $M^2$ from sum rule (\ref{sum rule m}). In the chosen work windows,
$M_{H}$ is calculated to be $4.38\pm0.09~\mbox{GeV}$.
In view of the uncertainty due to the variation of quark masses and
condensates, we have
$4.38\pm0.09^{+0.13}_{-0.07}~\mbox{GeV}$ (the
first error is resulted from the variation of $\sqrt{s_{0}}$
and $M^{2}$, and the second error reflects the uncertainty rooting in the variation of
QCD parameters) or briefly $4.38^{+0.22}_{-0.16}~\mbox{GeV}$
for $\Sigma_{c}\bar{D}$.

\begin{figure}[htb!]
\centerline{\epsfysize=7.0truecm\epsfbox{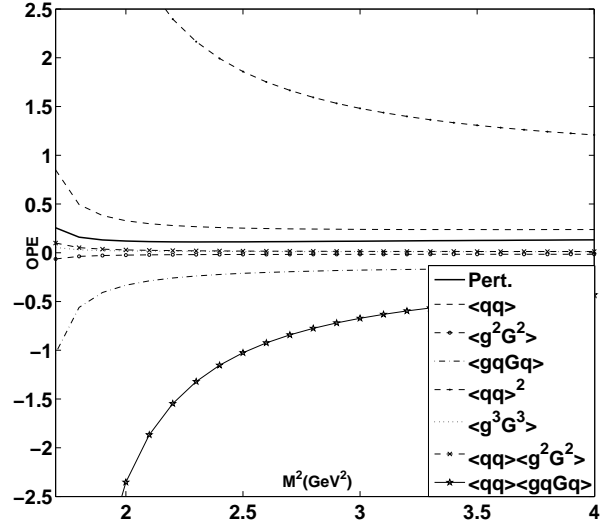}}
\caption{The relative contributions of various OPE as a function of $M^2$ in sum rule
(\ref{sumrule1}) for $\sqrt{s_{0}}=4.8~\mbox{GeV}$ for $\Sigma_{c}\bar{D}$.}
\end{figure}

\begin{figure}[htb!]
\centerline{\epsfysize=7.0truecm\epsfbox{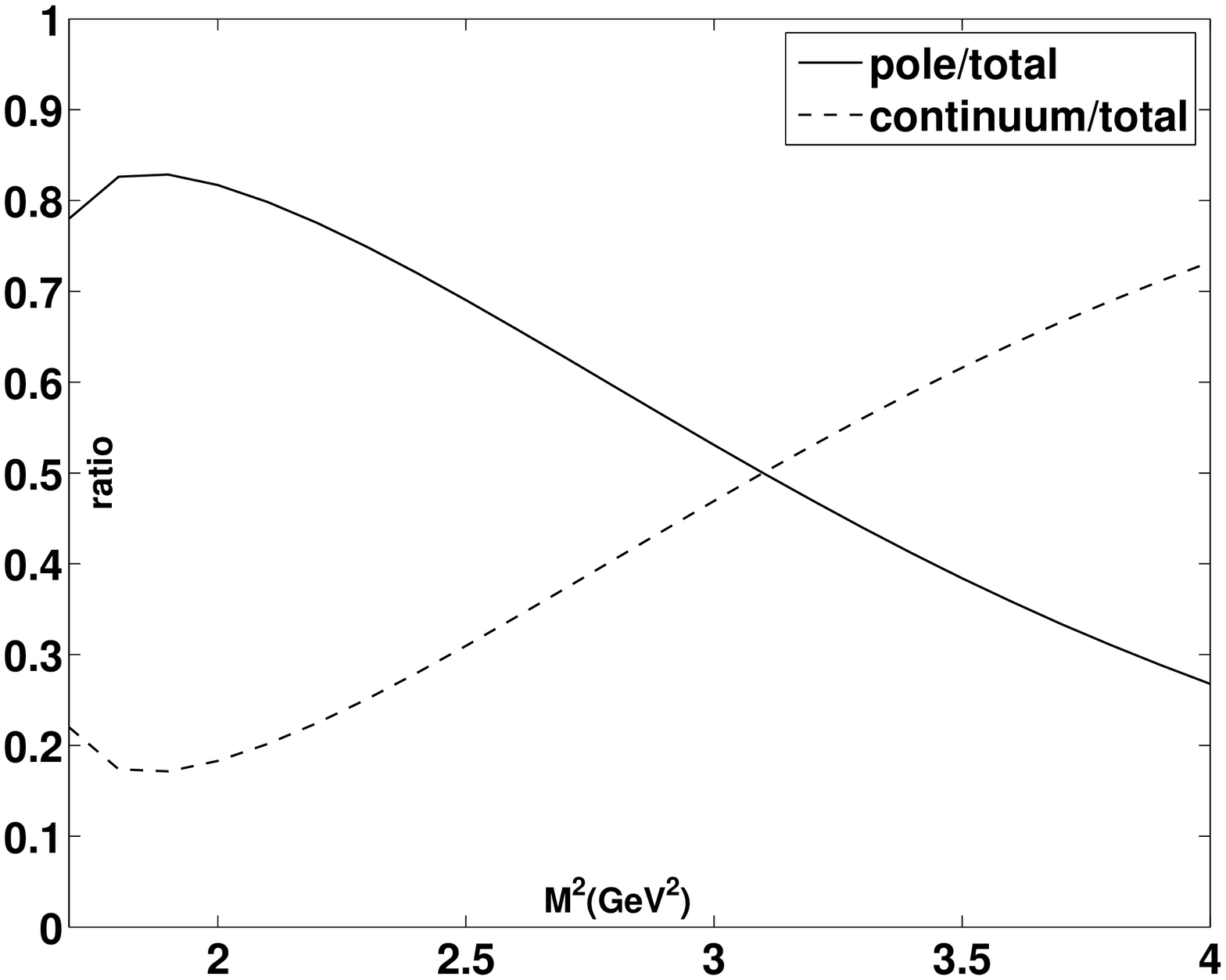}}
\caption{The phenomenological contribution in sum rule
(\ref{sumrule1}) for $\sqrt{s_{0}}=4.8~\mbox{GeV}$ for $\Sigma_{c}\bar{D}$.
The solid line is the relative pole contribution (the pole
contribution divided by the total, pole plus continuum contribution)
as a function of $M^2$ and the dashed line is the relative continuum
contribution.}
\end{figure}

\begin{figure}[htb!]
\centerline{\epsfysize=7.0truecm
\epsfbox{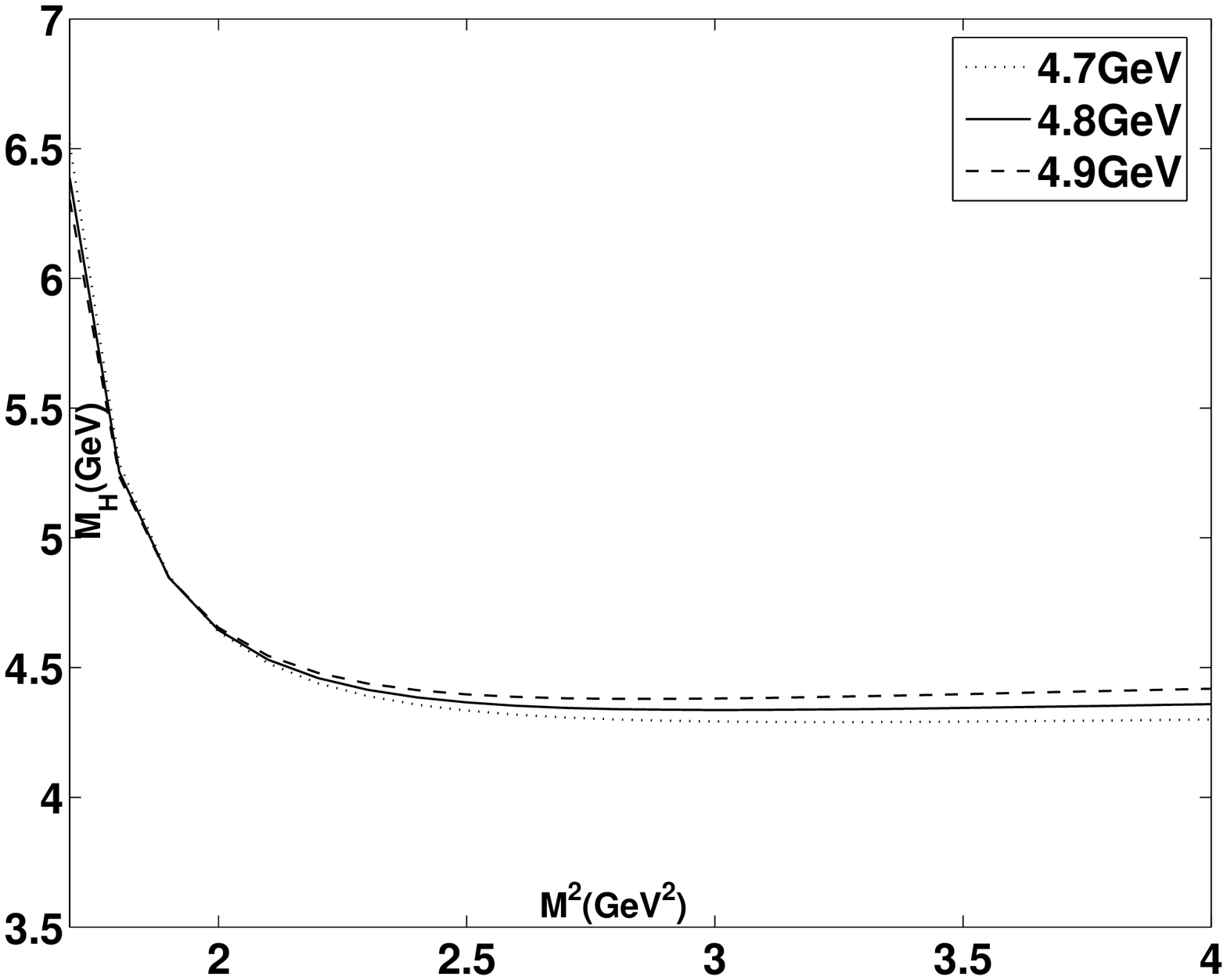}}\caption{
The mass of $\Sigma_{c}\bar{D}$ state as
a function of $M^2$ from sum rule (\ref{sum rule m}). The continuum
thresholds are taken as $\sqrt{s_0}=4.7\sim4.9~\mbox{GeV}$. The
ranges of $M^{2}$ are $2.2\sim2.9~\mbox{GeV}^{2}$ for
$\sqrt{s_0}=4.7~\mbox{GeV}$, $2.2\sim3.1~\mbox{GeV}^{2}$ for
$\sqrt{s_0}=4.8~\mbox{GeV}$, and $2.2\sim3.2~\mbox{GeV}^{2}$ for
$\sqrt{s_0}=4.9~\mbox{GeV}$.}
\end{figure}

In the end, combining the eventual results from both
(\ref{sum rule m}) and (\ref{sum rule q}), one could arrive at
 a conservative mass range $4.07\sim4.97~\mbox{GeV}$ for
the $\Sigma_{c}\bar{D}$ state,
which is consistent with the data of $P_{c}(4312)^{+}$ and could support
its explanation as a $\Sigma_{c}\bar{D}$ state.

\section{Summary}\label{sec4}
Motivated by LHCb's new discovery of
$P_{c}(4312)^{+}$, we study
that whether
$P_{c}(4312)^{+}$ could be a $\Sigma_{c}\bar{D}$ state
in QCD sum rules.
In order to insure the quality of sum rule analysis, contributions of condensates
up to dimension $8$ have been computed to test the OPE convergence.
We find that some condensates, i.e. the two-quark condensate,
the mixed condensate, the four-quark condensate, and the $\langle\bar{q}q\rangle\langle
g\bar{q}\sigma\cdot G q\rangle$ condensate are of importance to the OPE side.
Not bad, those main condensates could cancel each other out
to some extent, which brings that the OPE
convergence is still controllable.
By combining those results from two sum rules, we
finally obtain that a conservative mass range for $\Sigma_{c}\bar{D}$ is
$4.07\sim4.97~\mbox{GeV}$,
which is in agreement with the experimental value of $P_{c}(4312)^{+}$.
This result supports that $P_{c}(4312)^{+}$ could be explained as
a $\Sigma_{c}\bar{D}$ state.

In the future, one can expect that
further experimental observations
may shed more light on the nature of $P_{c}(4312)^{+}$ and
the inner structure of $P_{c}(4312)^{+}$
could be further revealed by continual efforts in both experiment and theory.

%%%%%%%%%%%%%%%%%%%%%%%%%%%%%%%%%%%%%%
\begin{acknowledgments}
This work was supported by the National
Natural Science Foundation of China under Contract
Nos. 11475258, 11105223, and 11675263, and by the project for excellent youth talents in
NUDT.
\end{acknowledgments}

%%%%%%%%%%%%%%%%%%%%%%%%%%%%%%%%%%%%%%%%%%%%%%%%%%%%


\begin{thebibliography}{99}

\bibitem{LHCb}R.~Aaij {\it et al.} (LHCb Collaboration), Phys. Rev. Lett. {\bf122}, 222001 (2019).

\bibitem{th-1}H.~X.~Chen, W.~Chen, and S.~L.~Zhu, arXiv:1903.11001 [hep-ph].

\bibitem{th-2}R.~Chen, Z.~F.~Sun, X.~Liu, and S.~L.~Zhu, Phys. Rev. D {\bf100}, 011502 (2019).

\bibitem{th-3}F.~K.~Guo, H.~J.~Jing, Ulf-G.~Mei{\ss}ner, and S.~Sakai, Phys. Rev. D {\bf99}, 091501 (2019).

\bibitem{th-4}M.~Z.~Liu, Y.~W.~Pan, F.~Z.~Peng, M.~S.~S\'{a}nchez,
L.~S.~Geng, A.~Hosaka, and M.~P.~Valderrama, Phys. Rev. Lett. {\bf122}, 242001 (2019).


\bibitem{th-5}J.~He, Eur. Phys. J. C {\bf79}, 393 (2019).


\bibitem{th-6}H.~X.~Huang, J.~He, and J.~L.~Ping, arXiv:1904.00221 [hep-ph].


\bibitem{th-7}A.~Ali and A.~Ya.~Parkhomenko, Phys. Lett. B {\bf793}, 365 (2019).

\bibitem{th-8}Y.~Shimizu, Y.~Yamaguchi, and M.~Harada, arXiv:1904.00587 [hep-ph].

\bibitem{th-9}Z.~H.~Guo and J.~A.~Oller, Phys. Lett. B {\bf793}, 144 (2019).

\bibitem{th-10}C.~J.~Xiao, Y.~Huang, Y.~B.~Dong, L.~S.~Geng, and D.~Y.~Chen, Phys. Rev. D {\bf100}, 014022 (2019).

\bibitem{th-11}C.~W.~Xiao, J.~Nieves, and E.~Oset, Phys. Rev. D {\bf100}, 014021 (2019).


\bibitem{th-12}X.~Cao and J.~P.~Dai, arXiv:1904.06015 [hep-ph].

\bibitem{th-13}H.~Mutuk, arXiv:1904.09756 [hep-ph].

\bibitem{th-14}X.~Z.~Weng, X.~L.~Chen, W.~Z.~Deng, and S.~L.~Zhu, Phys. Rev. D {\bf100}, 016014 (2019).


\bibitem{th-15}C.~Fern\'{a}ndez-Ram\'{\i}rez, A.~Pilloni, M.~Albaladejo, A.~Jackura,
V.~Mathieu, M.~Mikhasenko, J.~A.~Silva-Castro, and A.~P.~Szczepaniak (Joint Physics Analysis Center),
arXiv:1904.10021 [hep-ph].



\bibitem{th-16}R.~L.~Zhu, X.~J.~Liu, H.~X.~Huang, and C.~F.~Qiao, arXiv:1904.10285 [hep-ph].



\bibitem{others}L.~Maiani, A.~D.~Polosa, and V.~Riquer,
Phys. Lett. B {\bf749}, 289 (2015);
R.~F.~Lebed, Phys. Lett.
B {\bf749}, 454 (2015);
V.~V.~Anisovich, M.~A.~Matveev, J.~Nyiri, A.~V.~Sarantsev, and
A.~N.~Semenova, arXiv:1507.07652 [hep-ph];
G.~N.~Li, X.~G.~He, and M.~He, JHEP {\bf12}, 128 (2015);
R.~Ghosh, A.~Bhattacharya, and B.~Chakrabarti, Phys. Part. Nucl. Lett. {\bf14}, 550 (2017);
Z.~G.~Wang, Eur. Phys. J. C {\bf76}, 70 (2016);
R.~L.~Zhu and C.~F.~Qiao, Phys. Lett.
B {\bf756}, 259 (2016);
M.~Karliner and J.~L.~Rosner, Phys. Rev. Lett. {\bf115}, 122001 (2015);
R.~Chen, X.~Liu, X.~Q.~Li, and S.~L.~Zhu,
Phys. Rev. Lett. {\bf115}, 132002 (2015);
L.~Roca, J.~Nieves, and E.~Oset, Phys. Rev. D {\bf92}, 094003 (2015);
J.~He, Phys.
Lett. B {\bf753}, 547 (2016);
H.~X.~Huang, C.~R.~Deng, J.~L.~Ping, and F.~Wang,
Eur. Phys. J. C {\bf76}, 624 (2016);
F.~K.~Guo, Ulf-G.~Mei{\ss}ner, W.~Wang, and Z.~Yang, Phys. Rev. D {\bf92}, 071502 (2015);
Ulf-G.~Mei{\ss}ner and J.~A.~Oller,
Phys. Lett. B {\bf751}, 59 (2015);
X.~H.~Liu, Q.~Wang, and Q.~Zhao, Phys. Lett. B {\bf757}, 231 (2016);
M.~Mikhasenko, arXiv:1507.06552 [hep-ph]; M. Bayar, F. Aceti, F. K. Guo, and E. Oset, Phys. Rev.
D {\bf94}, 074039 (2016).




\bibitem{Penta-overview}Y.~R.~Liu, H.~X.~Chen, W.~Chen, X.~Liu, and S.~L.~Zhu, Prog. Part. Nucl. Phys. {\bf107}, 237 (2019).



\bibitem{svzsum}M.~A.~Shifman, A.~I.~Vainshtein, and V.~I.~Zakharov, Nucl. Phys. B {\bf147}, 385 (1979); B {\bf147}, 448 (1979);
 V.~A.~Novikov, M.~A.~Shifman, A.~I.~Vainshtein, and V.~I.~Zakharov, Fortschr. Phys. {\bf 32}, 585 (1984).




\bibitem{overview1}B.~L.~Ioffe, in The Spin Structure of The Nucleon, edited by
B.~Frois, V.~W.~Hughes, and N.~de Groot (World Scientific,
Singapore, 1997).




\bibitem{overview2}S.~Narison, Camb. Monogr. Part. Phys. Nucl. Phys. Cosmol. {\bf17}, 1
(2002).




\bibitem{overview3}P.~Colangelo and A.~Khodjamirian, in At the Frontier of
Particle Physics: Handbook of QCD, edited by M.~Shifman,
Boris Ioffe Festschrift Vol. 3 (World Scientific,
Singapore, 2001), pp. 1495-1576.


\bibitem{reinders}L.~J.~Reinders, H.~R.~Rubinstein, and S.~Yazaki, Phys. Rep. {\bf 127}, 1 (1985).


\bibitem{overview4}M.~Nielsen, F.~S.~Navarra, and S.~H.~Lee, Phys. Rep. {\bf497},
41 (2010).


\bibitem{Peta-SR1}H. X. Chen, W. Chen, X. Liu, T. G. Steele, and S. L. Zhu,
Phys. Rev. Lett. {\bf115}, 172001 (2015).


\bibitem{Peta-SR2}H. X. Chen, E. L. Cui, W. Chen, X. Liu, T. G. Steele,
and S. L. Zhu, Eur. Phys. J. C {\bf76}, 572 (2016).

\bibitem{Peta-SR3}K.~Azizi, Y.~Sarac, and H.~Sundu, Phys. Rev. D {\bf95},
094016 (2017).

\bibitem{Peta-SR4}Z.~G.~Wang, Int. J. Mod. Phys. A {\bf34}, 1950097 (2019).





\bibitem{Zs0}H.~X.~Chen, A.~Hosaka, and S.~L.~Zhu, Phys. Lett. B {\bf650},
369 (2007).

\bibitem{Zs1}Z.~G.~Wang, Nucl. Phys. A {\bf791}, 106 (2007).

\bibitem{Zs2}R.~D.~Matheus, F.~S.~Navarra, M.~Nielsen, and R.~Rodrigues
da Silva, Phys. Rev. D {\bf76}, 056005 (2007).


\bibitem{Zs}J.~R.~Zhang, L.~F.~Gan, and M.~Q.~Huang, Phys. Rev. D {\bf85}, 116007 (2012);
J.~R.~Zhang and G.~F.~Chen, Phys. Rev. D {\bf86}, 116006 (2012).


\bibitem{A} S.~Narison, Phys.
Rep. {\bf84}, 263 (1982); G.~Launer, S.~Narison, and R.~Tarrach, Z. Phys. C {\bf26}, 433 (1984);
S.~Narison, Riv. Nuovo Cimento {\bf10N2}, 1 (1987);
S.~Narison, World Sci. Lecture Notes Phys. {\bf26}, 1 (1989);
S.~Narison, Acta Phys. Polon. {\bf26}, 687 (1995);
S.~Narison, Phys. Lett. B {\bf673}, 30 (2009).


\bibitem{B}R.~Thomas, T.~Hilger, and B.~K\"{a}mpfer, Nucl. Phys. A {\bf795}, 19 (2007);
A.~G\'{o}mez Nicola, J.~R.~Pel\'{a}ez, and J.~Ruiz de Elvira, Phys. Rev. D {\bf82}, 074012 (2010);
H.~S.~Zong, D.~K.~He, F.~Y.~Hou, and W.~M.~Sun, Int. J. Mod. Phys. A {\bf23}, 1507 (2008);
S.~Narison, Phys. Lett. B {\bf707}, 259 (2012).



\bibitem{C}R.~Albuquerque, S.~Narison, A.~Rabemananjara, and D.~Rabetiarivony, Int. J. Mod. Phys. A {\bf31}, 1650093 (2016);
R.~Albuquerque, S.~Narison, F.~Fanomezana, A.~Rabemananjara, D.~Rabetiarivony, and G.~Randriamanatrika, Int. J. Mod. Phys. A {\bf31}, 1650196 (2016);
R.~Albuquerque, S.~Narison, D.~Rabetiarivony, and G.~Randriamanatrika, arXiv:1801.03073 [hep-ph].


\bibitem{baryon-current}B.~L.~Ioffe, Nucl. Phys. B {\bf188}, 317 (1981);
E.~V.~Shuryak, Nucl. Phys. B {\bf198}, 83 (1982).



\bibitem{Zhang}J.~R.~Zhang and M.~Q.~Huang, Phys. Rev. D {\bf77}, 094002 (2008);
JHEP {\bf1011}, 057 (2010); Phys. Rev. D {\bf83}, 036005 (2011).


\bibitem{Zhang1}J.~R.~Zhang, Phys. Rev. D {\bf87}, 076008 (2013); Phys. Rev. D {\bf89}, 096006 (2014).


\bibitem{Zhang2}J.~R.~Zhang, Phys. Lett. B {\bf789}, 432 (2019);
J.~R.~Zhang, J.~L.~Zou, and J.~Y.~Wu, Chin. Phys. C {\bf42}, 043101 (2018).


\bibitem{PDG}M.~Tanabashi {\it et al.} (Particle Data Group), Phys. Rev. D {\bf98},
030001 (2018).


\end{thebibliography}
\end{document}